\documentclass[twocolumn,english]{revtex4}
\usepackage[T1]{fontenc}
\usepackage[latin9]{inputenc}
\usepackage{graphicx}
\usepackage{esint}

\makeatletter
\@ifundefined{textcolor}{}
{%
 \definecolor{BLACK}{gray}{0}
 \definecolor{WHITE}{gray}{1}
 \definecolor{RED}{rgb}{1,0,0}
 \definecolor{GREEN}{rgb}{0,1,0}
 \definecolor{BLUE}{rgb}{0,0,1}
 \definecolor{CYAN}{cmyk}{1,0,0,0}
 \definecolor{MAGENTA}{cmyk}{0,1,0,0}
 \definecolor{YELLOW}{cmyk}{0,0,1,0}
 }

\makeatother

\usepackage{babel}

\begin{document}

\title{BKT phase transition in a 2d system with long range dipole-dipole
interaction.}

\author{P.O. Fedichev$^{1}$ and L.I. Men'shikov$^{2}$}

\affiliation{$^{1)}$Quantum Pharmaceuticals Ltd, Ul. Kosmonavta Volkova 6-606,
Moscow, Russian Federation}

\email{peter.fedichev@gmail.com}

\homepage{http://q-pharm.com}

\affiliation{$^{2)}$RRC Kurchatov Institute, Kurchatov Square 1, 123182, Moscow,
Russian Federation}
\begin{abstract}
We consider phase transitions in 2d XY-like systems with long range
dipole-dipole interactions and demonstrate that BKT-type phase transition
always occurs separating the ordered (ferroelectric) and the disordered
(paraelectric) phases. The low-temperature phase corresponds to a
thermal state with bound vortex-antivortex pairs characterized by
linear attraction at large distances. We show that bound vortex pairs
polarize and screen the vortex-antivortex interaction leaving only
the logarithmic attraction at sufficiently large separations between
the vortices. At higher temperatures the pairs dissociate and the
phase transition similar to BKT occurs, though at a larger temperature
than in a system without the dipole-dipole interaction.
\end{abstract}
\maketitle
Phase transitions and long range order in 2d systems are intimately
related with dynamics of topological excitations (vortices) with logarithmic
interactions: the transition occurs when vortex-antivortex pairs dissociate
at a certain finite temperature\citep{kosterlitz1973oma,berezinskii1971dlr,tsvelik2003qft}.
This conclusion is well established for a free system and does not
qualitatively change in a presence of a short range interaction. The
ground state of the system is not truly ordered (there can be no true
long range order) but superfluid at temperatures below the phase transition
point. The transition has been observed in numerous experimental systems,
such as, most recently, weakly interacting ultra-cold gases \citep{hadzibabic2006bkt}. 

There are situations though when the microscopic interactions are
long range, as in a case of a system with appreciable dipole-dipole
forces. The nature of the phase transition under the circumstances
has a long history \citep{maleev1976dft,pokrovskii1977dmi,bedanov1992lro,belobrov1983gss,brankov1987gsi,MaierShwabl,zimmerman1988pcd}.
Dipole-dipole interaction are shown to lead to formation of a true
long range order at small temperatures, so that the ground state of
a 2d system of magnetic dipoles is a ferromagnet \citep{maleev1976dft}.
The spontaneous polarization of the ground state has been also pointed
out in \citep{pokrovskii1977dmi}. Both works do not clear out the
nature of the phase transition, which has been studied first in \citep{feigelman1979},
where the correction to BKT temperature was obtained in the limit
when the dipole-dipole interaction is much smaller than the exchange
term, and hence the correction to the transition temperature is small.
The author made an important observation though: the interaction between
the vortices was found to be linear at sufficiently large distances
(see \citep{MaierShwabl} for a recent discussion). Ferromagnetic
ordering was studied in various model dipole systems theoretically
in \citep{bedanov1992lro,brankov1987gsi,belobrov1983gss} and experimentally
in \citep{zimmerman1988pcd}. Arguably, dipole-dipole forces between
water molecules on hydrophobic surfaces lead to macroscopic ordering
of the molecular dipoles such as creation of macroscopic hydrogen-bond
networks in biological systems \citep{fedichev-2006}. 

In spite of vast volume of the research, the nature of the disordering
phase transition appears to be somewhat controversial. There are claims
that the long range dipole-dipole forces change the physics of the
phase transition entirely transforming BKT vortex pairs dissociation
transition in a free system to a confinement transition similar to
that in quark-gluon plasma \citep{MaierShwabl}. Note, that transition
temperature is predicted to be four times higher than BKT temperature.
Below we perform a systematic study of dipole-dipole interaction influence
in 2d systems at finite temperatures. We show that the interaction
leads to ferromagnetic ordering at low temperatures without contradiction
to Mermin-Peierls theorem \citep{Mermin} and in a full accordance
to the earlier statements. The phase transition itself turns out,
as in BKT case, to be associated with dissociation of vortices. We
calculate the transition temperature, $T_{C}$, as a function of the
interactions parameters. We show that the linear interaction between
the vortices does change the transition temperature, though the vortex
gas polarization screens the long range linear potential and transforms
it to a logarithmic interaction. Therefore, in a system with dipole-dipole
interaction the phase transition is essentially BKT, though the transition
temperature itself has a complicated dependence on the model parameters. 

At last we apply the developed model to water-solute boundaries using
the phenomenological vector model of polar liquids \citep{fedichev-2006}.
We show that spontaneous polarization of molecular dipoles next to
hydrophobic boundaries may occur and a ferroelectric liquid film may
form. The vector model naturally describes topological excitations
on the solute boundaries. Since the molecular dipoles interact electrostatically
at large distances, the dissociation of the vortex pairs can be associated
with disappearance of hydrogen bond networks in the course of order-disorder
phase transition in the hydration water layer \citep{Oleinikova1,Oleinikova2,Oleinikova3}.

Consider a plane layer of a thickness $\lambda$ composed of interacting
dipoles. The unit vector ${\bf S}(\mathbf{r})$ is taken parallel
to the dipole moment of a molecule residing at a point $\mathbf{r}$
and is characterized by the orientation angle $\theta({\bf \textbf{r}})$:
$\mathbf{S}=(\cos\theta(\textbf{r}),\sin\theta(\textbf{r}))$. The
Hamiltonian of the interacting system consists of the two parts: \begin{equation}
G_{S}=G_{H}+G_{dd},\label{eq: the Hamiltonian}\end{equation}
where the short ranged gradient term gives the energy of a free system\begin{equation}
G_{H}=\frac{1}{2}M\int_{\Gamma}df\left(\nabla\theta\right)^{2},\label{eq: G H}\end{equation}
and the long range dipole-dipole interaction is represented by the
following model term: \[
G_{dd}=\frac{1}{2}K\int_{\Gamma}dfdf^{\prime}\frac{\left(\nabla\cdot\mathbf{S}\right)(\nabla^{\prime}\cdot\mathbf{S}^{\prime})}{\left|\mathbf{r}-\mathbf{r}^{\prime}\right|},\]
where $df$ is the element of the surface $\Gamma$, whereas the constants
$M$ and $K$ characterize the strength of the interactions. The Hamiltonian
has been studied in the limit $M\gg\lambda K$ in \citep{feigelman1979,maleev1976dft}. 

Consider the case of an arbitrary relation between parameters $\lambda K$
and $M$. At small temperatures $T$ the dipole system is ordered,
i.e. all the dipoles point at the same direction, for example, along
the $x$-axis ($\theta(\mathbf{r})=0$). To prove that let us consider
the correlation function of the dipole orientation fluctuations: $K_{2}\left(\mathbf{r}\right)=\langle\theta\left(0\right)\theta\left(\mathbf{r}\right)\rangle$,
$\left|\theta\right|\ll1$. Keeping quadratic terms in $\theta$ only
and linearizing the Hamiltonian we find that \[
G_{S}\approx\frac{1}{2}M\int_{\Gamma}df\left(\nabla\theta\right)^{2}+\frac{1}{2}K\int_{\Gamma}dfdf^{\prime}\frac{\theta_{y}\left(\mathbf{r}\right)\theta_{y}\left(\mathbf{r}^{\prime}\right)}{\left|\mathbf{r}-\mathbf{r}^{\prime}\right|}\]
so that the correlation function takes the form \begin{equation}
K_{2}\left(\mathbf{r}\right)=T\int d^{2}k\frac{\exp\left(i\mathbf{kr}\right)}{Mk^{2}+2\pi Kk\sin^{2}\alpha},\label{eq: K on r}\end{equation}
where $\alpha$ is the angle between wave vector $\mathbf{k}$ and
$x$-axes. The integral in the r.h.s. converges at small values of
$k$, therefore $K_{2}(0)$ is finite and hence the ordered state
is thermally stable. In fact the appearance of the linear term in
the integrand denominator resolves the contradiction with Mermin-Peierls
theorem \citep{Mermin}. It may even seem that the correlation function
integral converges solely due to the finite value of the dipole dipole
interaction strength ($K\neq0$). The conclusion is in fact wrong:
as it is clear from Eq. (\ref{eq: K on r}), at $M\rightarrow0$ the
integral diverges at $\alpha\rightarrow0$. It means that both terms
in Eq. (\ref{eq: the Hamiltonian}) denominator are both equally important.
Mathematically this analysis is in full agreement with a well known
feature of classic electrostatics, the Earnshaw theorem: a system
of classical charges interacting with electrostatic forces only can
have no stable state \citep{jackson1999ce}. 

To elucidate the nature of the ordered state let us consider the correlation
function at large distances $r$. The main contribution to the integral
in Eq. (\ref{eq: K on r}) comes from the two separate regions: $\left|\alpha\right|\ll1$
and $\left|\alpha-\pi\right|\ll1$, so that \begin{equation}
K_{2}\left(\mathbf{r}\right)\approx\frac{4T}{M\sqrt{\pi}}\int_{0}^{\infty}d\tau\cos\left(x\tau^{2}\right)\exp\left(-\frac{\left|y\right|\tau^{3}}{\sqrt{\gamma}}\right)\label{eq: K on r analytical}\end{equation}
with $\gamma=2\pi K/M$. The calculation in Eq.(\ref{eq: K on r analytical})
simplifies in the two following limiting cases:\[
K_{2}\approx\frac{\sqrt{2}T}{M\sqrt{\left|x\right|}},\;\left|y\right|\ll\sqrt{\gamma}\left|x\right|^{3/2},\]
\[
K_{2}\approx\frac{4T}{3M\sqrt{\pi}\left|y\right|^{1/3}}\Gamma\left(\frac{1}{3}\right),\;\left|y\right|\gg\sqrt{\gamma}\left|x\right|^{3/2}.\]
The correlations at large temperatures are addressed in \citep{MaierShwabl}
and turn out to decay exponentially. The crossover between the power
laws found here for small $T$ and the exponential decay at large
$T$ suggests a phase transition at some intermediate temperature
$T_{C}$.%
\begin{figure}
\includegraphics[width=0.9\columnwidth]{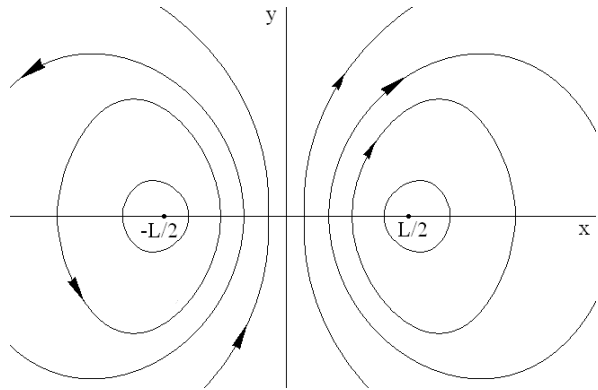}

\caption{Polarization configuration $\mathbf{S}(x,y)$ corresponding to a vortex-antivortex
pair (VAP).\label{fig: Vortex antivortex pair}}

\end{figure}

To establish the transition temperature we apply a set of familiar
arguments (see e.g. \citep{tsvelik2003qft} for further explanations).
Thermodynamics of a 2d system can be mapped to thermodynamics of a
gas of interacting vortex-antivortex pairs (VAPs) \citep{minnhagen1987tdc}.
Consider first the small temperatures limit:$T<T_{C}$, $T_{C}-T\ll T_{C}$.
The thermal state in this case can be viewed as a gas of bound VAPs.
Since the considered temperature is still below the transition point,
VAPs approach the dissociation limit and the pairs with large distances
$r$ between the vortices dominate. At such separations the vortex-antivortex
interaction is described by the linear term and the energy of the
system can be approximated as:\begin{equation}
F\approx NE_{0}+\frac{1}{2}K\sum_{i,\, j}q_{i}q_{j}\left(-r_{ij}\right)=\frac{1}{2}\sum_{i}q_{i}\Phi\left(\mathbf{r}_{i}\right),\label{eq: Simplified VAP interaction}\end{equation}
where $N$ is the number of the VAPs and $q_{j}=\pm1$ are the topological
charges associated with the vortices. The {}``rest energy'', $E_{0}$,
is the energy associated with a pair of a minimum possible separation
$r\sim\lambda$, where $\lambda$ is the size of the vortex core.
The latter quantity is is small, model dependent and depends on microscopic
details of the underlying physical system (such as thickness of the
layer). The (quasi-electric) potential $\Phi\left(\mathbf{r}\right)$
is introduced in analogy with classic electrostatics and amounts to
the energy associated with the interaction of a given vortex (charge)
with all other vortices in the system:\begin{equation}
\Phi\left(\mathbf{r}\right)=-K\sum_{j}q_{j}\left|\mathbf{r}-\mathbf{r}_{j}\right|=-K\int\left|\mathbf{r}-\mathbf{r}^{\prime}\right|\rho\left(\mathbf{r}^{\prime}\right)d^{2}\mathbf{r}^{\prime},\label{eq: Expression for Phi}\end{equation}
where\begin{equation}
\rho\left(\mathbf{r}\right)=\sum_{j}q_{j}\delta^{\left(2\right)}\left(\mathbf{r}-\mathbf{r}_{j}\right)\label{eq: Expression for vortex charge density}\end{equation}
is the vortex charge density. Note, that according to its definition
(\ref{eq: Expression for Phi}), the potential $\Phi\left(\mathbf{r}\right)$
satisfies the analogue of Poisson equation\[
\hat{L}_{\mathbf{r}}\Phi\left(\mathbf{r}\right)=K\rho\left(\mathbf{r}\right),\]
where the linear operator $\hat{L}_{\mathbf{r}}$ is defined so that
\begin{equation}
\hat{L}_{\mathbf{r}}\left(-\left|\mathbf{r}-\mathbf{r}^{\prime}\right|\right)=\delta^{\left(2\right)}\left(\mathbf{r}-\mathbf{r}^{\prime}\right).\label{eq: Definition of L on r}\end{equation}
Let us follow the electrostatic analogy even further: since the energy
of a charge (vortex) $q$ placed in the external potential $\Phi$
is $U=q\Phi,$ then the force, acting on the charge (vortex) is $\mathbf{F}=q\mathbf{E}$,
where the quasi-electric field vector $\mathbf{E}=-\nabla\Phi$ is
associated with the potential $\Phi$ in a normal way.

To find out the energy of a VAP in a self-consistent {}``electric''
field of all other pairs, let us define the {}``dipole moment''
of a pair according to \begin{equation}
\mathbf{d}=\sum_{j}q_{j}\mathbf{r}_{j}=q_{+}\mathbf{r}_{+}+q_{-}\mathbf{r}_{-}\equiv\mathbf{r},\label{eq: Definition of dipole moment}\end{equation}
 where $\mathbf{r}=\mathbf{r}_{+}-\mathbf{r}_{-}$. Then, the energy
of the pair is \begin{equation}
U=q_{+}\Phi\left(\mathbf{r}_{+}\right)+q_{-}\Phi\left(\mathbf{r}_{-}\right)\approx-\mathbf{d\cdot E}.\label{eq: energy of dipole on external field}\end{equation}
Let us pursue the analogy and calculate first the polarizability $\alpha_{P}$
of a single VAP. The dipole moment of a pair in a weak external field
$\mathbf{E}$ is given by a standard relation\begin{equation}
\left\langle \mathbf{d}\right\rangle =\frac{\int df\mathbf{r}\exp\left(\frac{\mathbf{r\cdot E}}{T}-\frac{K}{T}r\right)}{\int df\exp\left(\frac{\mathbf{r\cdot E}}{T}-\frac{K}{T}r\right)}\approx\alpha_{P}\mathbf{E},\label{eq: Average VAP dipole moment}\end{equation}
where $\alpha_{P}=3T/K^{2}$ is nothing else but the pair polarizability.

At the transition temperature, $T=T_{C}$, VAPs begin to dissociate.
It means that at $T_{C}-T\ll T_{C}$ only a small fraction of the
pairs are very large and close to dissociation. For this reason it
is possible to neglect the interactions between the largest VAPs and
calculate the energy of single large pair approaching its dissociation
limit in a cloud of comparatively small bound VAPs. As we have demonstrated
above, the bound pairs are polarizable and therefore the field of
a charge is screened by the polarization of VAPs gas, thus influencing
the potential energy of a large VAP. To find out how exactly, let
us consider the effect of shielding of a probe point charge $Q$ or
the charge density $\rho_{Q}=Q\delta^{\left(2\right)}\left(\mathbf{r}\right)$
placed somewhere at the origin. The complete {}``electrostatic''
potential $\Phi\left(\mathbf{r}\right)$ is produced both by the probe
charge and the polarization charges of VAPs. The density $\rho_{P}$
of the polarization charges is given by usual expressions following
from Eq. (\ref{eq: Definition of dipole moment}): $\rho_{P}=-\nabla\cdot\mathbf{P},$
where $\mathbf{P}$ is polarization of the vortex gas:\begin{equation}
\mathbf{P}=n_{P}\left\langle \mathbf{d}\right\rangle =\chi\mathbf{E}=-\chi\nabla\Phi,\label{eq: Polarization vector of VAP gas}\end{equation}
with $\chi=\alpha_{P}n_{P}$ and $n_{P}$ being the {}``dielectric
susceptibility'' of the gas of VAPs and the concentration of the
pairs, correspondingly. 

Combining Eqs. (\ref{eq: Definition of L on r}), (\ref{eq: Polarization vector of VAP gas})
and (\ref{eq: Expression for Phi}) we obtain the complete equation
for $\Phi\left(\mathbf{r}\right)$:\begin{equation}
\hat{L}_{\mathbf{r}}\Phi\left(\mathbf{r}\right)=K\left(\rho_{Q}+\rho_{P}\right)=KQ\delta^{\left(2\right)}\left(\mathbf{r}\right)+K\chi\triangle\Phi\left(\mathbf{r}\right).\label{eq: Equation for phi}\end{equation}
Since the Fourier component of the operator $\bar{L}_{\mathbf{r}}$
is $L_{\mathbf{k}}=k^{3}/2\pi$, we find that \begin{equation}
\Phi(\mathbf{r})=QK\int\frac{d^{2}k}{\left(2\pi\right)^{2}}\frac{\exp\left(i\mathbf{kr}\right)}{\frac{k^{3}}{2\pi}+K\chi k^{2}}.\label{eq: Expression for F}\end{equation}
At large distances, $r\gg r_{0}=1/2\pi K\chi$, most contribution
to the integral comes from small values of $k$ where $k^{3}$ term
in the denominator is negligible. Therefore the potential of the {}``charge''
at large distances is logarithmic: \begin{equation}
\Phi(\mathbf{r})=-\frac{Q}{2\pi\chi}\log\left(\frac{r}{C_{1}}\right),\quad r\gg r_{0},\label{eq: r gg r 0}\end{equation}
where $C_{1}\sim r_{0}$. In the opposite limit, $r\ll r_{0}$, the
potential is linear: $\Phi(\mathbf{r})\approx-Kr,\quad r\ll r_{0}.$
Below we propose a simple expression interpolating between the two
results:\begin{equation}
\Phi(\mathbf{r})\approx-\frac{Q}{2\pi\chi}\log\left(1+\frac{r}{r_{0}}\right),\label{eq: F versus r}\end{equation}
so that the energy of a large pair of a size $R$ is given by\begin{equation}
E_{VAP}(R)\approx2E_{0}+\frac{1}{2\pi\chi}\log\left(1+\frac{R}{r_{0}}\right).\label{eq: Modified VAP energy}\end{equation}
We note that although the dipole-dipole interactions do change the
interactions between the vortices at small distances, the polarization
of the VAPs destroys the linear attraction at large separations between
the vortices and hence the phase transition associated with the dissociation
of the pairs is qualitatively very similar to BKT transition in a
free system.

Standard calculation of BKT temperature for a vortex gas with interaction
(\ref{eq: Modified VAP energy}) gives the following implicit equation
for the transition temperature $T_{C}$ %
\begin{figure}
\includegraphics[width=0.9\columnwidth]{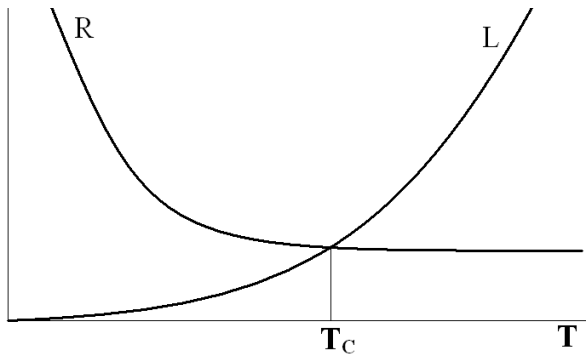}

\caption{Graphical solution of implicit equation (\ref{eq: Implicit equation for T C})
for the BKT-transition temperature $T_{C}$. By $R$ and $L$ graphics
of the r.h.s. and l.h.s. are shown, correspondingly.\label{fig: grafical solution of implicit equation}}

\end{figure}
 \begin{equation}
T_{C}^{2}=\frac{K^{2}}{12\pi n_{P}(T_{C})}.\label{eq: Final formula for transition temperature}\end{equation}
The density of $n_{P}$ at the transition temperature can be calculated
from the following argument. At $T\approx T_{c}$ the VAPs just start
to dissociate and therefore the size of a typical pair is small and
the interaction between the vortices is still linear and unscreened
and \begin{equation}
n_{P}(T)=2\pi\exp\left(-\frac{2E_{0}}{T}\right)\frac{T^{2}}{\lambda^{4}K^{2}}.\label{eq: Final result for n P}\end{equation}
Therefore, the Eq. (\ref{eq: Final formula for transition temperature})
for the transition temperature takes form\begin{equation}
2\pi\sqrt{6}\left(\frac{T_{C}}{\lambda K}\right)^{2}=\exp\left(\frac{E_{0}}{T_{C}}\right)\label{eq: Implicit equation for T C}\end{equation}
The graphical solution of the equation is demonstrated on Fig. \ref{fig: grafical solution of implicit equation}.
As it is clear from the graph, the solution of the equation always
exists and is unique. 

Let us assume first that the dipole-dipole interactions is small,
i.e. $\lambda K\ll M$. Since the energy of a minimal vortex-antivortex
pair $E_{0}\sim M$, the transition temperature $T_{c}\sim M$, which
coincides with the standard BKT result ($T_{BKT}=\pi M/2$). This
is precisely the limit considered in \citep{feigelman1979}. In the
opposite limit, $M\ll\lambda K$, the exponent in the r.h.s. of Eq.
(\ref{eq: Implicit equation for T C}) is approximately $1$ and hence
$T_{C}\sim\lambda K\gg T_{BKT}$. The phase transition temperature
is indeed larger than BKT temperature, in agreement with \citep{MaierShwabl},
although our analysis suggests that $T_{C}$ can be arbitrary larger
than $T_{BKT}$.

Layers of water molecules on hydrophobic surfaces such as biological
membranes or large bio-molecules are another interesting example of
2d systems with dipole-dipole interactions. The orientations of molecules
next to a macromolecular surface depend on interplay of long-range
interactions of the molecular dipoles and the short-range hydrogen
bonds between the adjacent molecules. Since an energy of uncompensated
hydrogen bond is large compared to the temperature, no water molecules
can point in the direction of a hydrophobic surface, and thus all
the water molecules next to the surface have their dipole moments
parallel to the surface. Macroscopic polarization of the molecules
$\mathbf{s}\left(\mathbf{r}\right)=\langle\mathbf{S}\left(\mathbf{r}\right)\rangle$
vanishes quickly in the direction of the liquid bulk. Therefore a
water layer next to a hydrophobic boundary can be studied with a help
of a model Hamiltonian (\ref{eq: the Hamiltonian}) corresponding
to a 2d of interacting dipoles (see \citep{fedichev-2006} and references
therein). According to the findings of the current work the layer
of water molecules is completely polarized, $s\approx s_{\parallel}$,
at very low temperatures. Microscopic parameters of water are such
that $M\sim\lambda K$ and therefore the macroscopic polarization
and molecules orientation ordering disappears at temperature \begin{equation}
T_{C}\sim M\sim E_{H}s_{\Vert}^{2},\label{eq: Estimation for T C}\end{equation}
where $E_{H}\approx2500^{0}K$ is the characteristic energy of a hydrogen
bond. The expression suggests that the transition temperature corresponding
a completely hydrophobic surface ($s_{\parallel}\approx1$) is always
very large, the water molecules are ordered at all realistic temperatures
and thus there always exists a macroscopic hydrogen bonds network
on the surface of the body. If a surface is partly hydrophillic, i.e.
hydrogen bonds donors or acceptors , such as charges, $s_{\parallel}<1$
and already at $s_{\parallel}=0.3$ the transition may occur already
at room temperatures. The evidence of such transitions is observed
in molecular dynamics calculations \citep{Oleinikova1,Oleinikova2,Oleinikova3}. 

The work was supported by Quantum Pharmaceuticals. Phenomenological
water models are used in Quantum everywhere to compute solvation energies
and protein-drug interactions in aqueous environment.

\bibliographystyle{prsty}
\bibliography{../Qrefs}

\end{document}